\documentclass[aps,floats,amssymb,twocolumn,prd]{revtex4}
\usepackage{calc}
\usepackage{psfrag}
\usepackage{graphicx}
\usepackage{color}

\def\ds{\stackrel{\star}{,}}
\def\tr{{\rm Tr}}
\def\pa{\partial}
\def\nn{\nonumber}
\def\a{\alpha}
\def\b{\beta}

\def\d{\delta}
\def\g{\gamma}

\def\k{\kappa}
\def\l{\lambda}
\def\m{\mu}
\def\n{\nu}

\def\r{\rho}                    %       \varrho
\def\s{\sigma}                  %       \varsigma

\def\L{\Lambda}

\begin{document}

\title{Noncommutative gravity and the relevance of the $\theta$-constant deformation}

\author{M. Dimitrijevi\'c \'Ciri\'c, B. Nikoli\'c and V. Radovanovi\'c}
\affiliation{Faculty of Physics, University of
Belgrade, Serbia}

\begin{abstract}

In this Letter we construct the noncommutative (NC) gravity model on the $\theta$-constant NC
space-time. We start from the NC $SO(2,3)_\star$ gauge theory and use the enveloping algebra
approach and the Seiberg-Witten map to construct the effective NC gravity action. The action and
the equations of motion are expanded up to second order in the deformation parameter. The equations
of motion show that the noncommutativity plays a role of a source 
for the curvature
and/or torsion. Finally, we calculate the NC corrections to the Minkowski space-time and we show
that in the presence of noncommutativity Minkowski space-time becomes curved, but remains
torsion-free. The breaking of diffeomorphism invariance is understood in terms 
of the preferred
coordinate system. We show that the coordinates we are using are the Fermi normal coordinates. This
suggests that the natural coordinate system in which one should study NC gravity is given by the
Fermi normal coordinates.

\end{abstract}

\maketitle

{\it NC gravity, general:} Formulated at the beginning of the XX century, General Relativity and
Quantum Mechanics continue to
remarkably well describe physics at different scales: the scale of planets and 
the Solar system
and the scale of atoms respectively. Predictions of both
theories have been confirmed by many experiments, the most recent being the detection of
gravitation waves \cite{GRWaves}. However, a theory that could unite these two theories and provide
a description of gravity at quantum scales is still missing. There are different proposals in
the literature: string theory, quantum loop gravity, noncommutative (NC) geometry, dynamical
triangulation,\dots. None of these proposals is complete, mostly due to the lack of
experimental tests.

NC gravity theories rely on the notion of noncommutative space-time and/or noncommutative geometry.
NC geometry provides a generalization of description of space-time by smooth manifolds.
Introducing NC relations between coordinates naturally leads to uncertainty 
relations
between coordinates \cite{DFR} and introduces a "foamy, grainy" structures instead of the usual
smooth manifolds. In this way one hopes to solve the problem of divergences in Quantum Field
Theories and the problem of singularities in General Relativity.

During the last fifteen years, different models of NC gravity theories were considered. We
just mention a few different approaches: NC gravity via the twist approach \cite{TwistApproach},
emergent NC gravity via the matrix models \cite{EmGravityApproach}, fuzzy space gravity models and
DFR models \cite{OtherApproaches} and NC gravity via the NC gauge theory and 
the Seiberg-Witten
(SW) map \cite{SWmapApproach, PL09}.

In this Letter, we study one particular model of NC gravity as a NC gauge theory. We work with
the "canonical" or "$\theta$-constant" noncommutativity. The main advantage of 
this NC space-time
is that, due to the constant noncommutativity, it is relatively easy to study various physics
problems. On the other hand, by introducing a constant NC parameter we 
explicitly break the
diffeomorphism symmetry.
Therefore, we need to understand if this symmetry breaking has some physical explanation.  The gauge
group is chosen to be the 
$SO(2,3)$ group. Motivated by different $f(R)$, $f(T)$ and other modified gravity models we study
the SW map expansion of our model and obtain correction terms that can be compared with the existing
terms in the modified gravity models. In our model the relations between different correction terms
are not arbitrary but are fixed by the SW map expansion. Calculating NC gravity equations of
motion, we show that noncommutativity plays a role of a source for the curvature and torsion. That
is, given a flat/torsion-free space-time, noncommutativity induces nonzero curvature/torsion on
this space-time. This result is not completely new, it was also discussed in \cite{MajaJohn} in a
different approach to NC gravity. Especially, starting from the Minkowski 
space-time as a solution of
commutative
vacuum Einstein equations, the corrections induced by our NC gravity model lead to a space-time with
a constant scalar curvature, (A)dS-like space-time. 

As we mentioned earlier, our model breaks the commutative diffeomorphism symmetry. This symmetry
breaking can be understood in terms
of a preferred coordinate system we are using.
Looking more carefully, we find that this particular coordinate system is given by the Fermi normal
coordinates. These are coordinates of an observer moving along a geodesic
in curved space-time. In particular,  this is the observer who measures the 
noncommutativity to
be constant. This provides an explanation of the
$\theta$-constant noncommutativity (deformation). It also suggests that the Fermi normal
coordinates are the preferred coordinate system in which one should study NC 
gravity effects.

\vspace*{0.2cm}
{\it Commutative model:} We start from a commutative gravity models based on the $SO(2,3)$ gauge
symmetry. Models of this type appeared in the 1970ies \cite{stelle-west} and were used in the
construction
of supergravity theories. The gauge field and the field strength tensor are defined as
\begin{eqnarray}
\omega_\mu &=& \frac{1}{2}\omega_\mu^{AB}M_{AB} = \frac{1}{4}\omega^{ab}\sigma_{ab} +
\frac{1}{2l}e_\mu^a\gamma_a \, ,\label{GaugeField}\\
F_{\mu\nu} &=& \frac{1}{2}F_{\mu\nu}^{AB}M_{AB} = \partial_\mu\omega_\nu - \partial_\nu\omega_\mu
-i[\omega_\mu,\omega_\nu] \label{FSTensor}\\
&=& \frac{1}{4}\Big( R_{\m\n}^{ab} - \frac{1}{l^2}(e_\m^a e_\n^b - e_\m^b e_\n^a)\Big)\sigma_{ab}
+\frac{1}{2l}T_{\m\n}^a\gamma_a \, ,\nn
\end{eqnarray}
with $e_\mu^{\ a} = l\omega_\mu^{a5}$ and $T_{\m\n}^a = lF_{\m\n}^{a5}$. The generators $M_{AB}$
close the $SO(2,3)$ algebra
\begin{eqnarray}
&&[M_{AB}, M_{CD}] = i(\eta_{AD}M_{BC}+\eta_{BC}M_{AD}\nn\\
&& \hspace*{2.8cm}-\eta_{AC}M_{BD}-\eta_{BD}M_{AC}) \,
,\label{AdSalgebra}\\
&&\eta_{AB} = (+,-,-,-,+),\quad A,B=0,1,2,3,5\ .\nn
\end{eqnarray} 
The quantities in (\ref{FSTensor}) are given by
\begin{eqnarray}
R_{\m\n}^{ab} &=& \pa_\m\omega_\n^{ab} - \pa_\n\omega_\m^{ab}
+ \omega_{\m c}^a\omega_\n^{cb} - \omega_{\m c}^b\omega_\n^{ca}, \nonumber\\
T_{\m\n}^a &=&  \nabla_\m e^{\ a}_\n - \nabla_\n e^{\ a}_\m  , \nn\\
\nabla_\m e^{\ a}_\n &=& \partial_\mu e^{\ a}_\n + \omega^a_{\mu b}e^{\ b}_\n \, .
\nonumber
\end{eqnarray}
More details about the $SO(2,3)$ algebra and the calculations that follow can be 
found in
\cite{UsInPreparation}.
The commutative action that we are going to generalize to the NC setting is a sum of
three terms:
\begin{eqnarray}
&&\hspace*{-5mm}S = c_1S_1+c_2S_2+c_3S_3\nn\\
&&\hspace*{-5mm}S_1 = \frac{il}{64\pi G_N}\tr \int{\rm d}^4x \epsilon^{\mu\nu\rho\sigma}
F_{\mu\nu} F_{\rho\sigma}\phi \, , \label{KomDejstvo_S_1}\\
&&\hspace*{-5mm}S_2 = \frac{1}{64 \pi G_{N}l}\tr\int {\rm d}^{4}x\epsilon^{\mu \nu \rho
\sigma}F_{\mu 
\nu}D_{\rho}\phi D_{\sigma}\phi\phi , \label{KomDejstvo_S_2}\\
&&\hspace*{-5mm}S_3 = \frac{-i}{128 \pi G_{N}l}\tr\int {\rm d}^{4}x\epsilon^{\mu \nu \rho 
\sigma}D_{\m}\phi D_{\n}\phi D_{\rho}\phi D_{\sigma}\phi\phi ,\label{KomDejstvo_S_3} 
\end{eqnarray}
with an additional scalar field $\phi=\Phi^A\Gamma_A,\>\>\Gamma^A=(i\gamma_a\gamma_5,
\gamma_5)$\footnote{Matrices
$\gamma_a$ are the usual four dimensional Dirac gamma matrices.} transforming in 
the adjoint
representation of $SO(2,3)$ and $D_\mu\phi = \partial_\mu\phi -i[\omega_\mu, \phi]$. The field
$\phi$ is used to break the $SO(2,3)$ gauge symmetry of the action to the $SO(1,3)$ (local Lorentz)
gauge symmetry. The action (\ref{KomDejstvo_S_1}) was introduces by Stelle and West and also
analyzed by Towsend and by MacDowell and Mansouri in their papers \cite{stelle-west}. Actions
(\ref{KomDejstvo_S_2}) and (\ref{KomDejstvo_S_3}) were introduced by Wilczek in \cite{Wilczek} as
different possibilities to write $SO(2,3)$ gauge invariant models.

The symmetry breaking is done by choosing a particular form of the field $\phi$, namely
$\phi^a=0,\phi^5=l$. The symmetry breaking can also be spontaneous, for more details see
\cite{Wilczek}. After the symmetry breaking the resulting commutative action is written as
\begin{eqnarray}
S &=& c_1S_1+c_2S_2+c_3S_3\nn\\
&=&-\frac{1}{16\pi G_{N}}\int 
d^{4}x\Big(c_1\frac{l^2}{16}\epsilon^{\m\n\r\s}
\epsilon_{abcd}R_{\m\n}^{\ ab}R_{\r\s}^{\ cd} \nn\\
&& +\sqrt{-g}\big( (c_1 + c_2)R -\frac{6}{l^2}(c_1+ 2c_2 + 2c_3)\big)
\Big), \label{KomDejstvo} 
\end{eqnarray}
with $\sqrt{-g}=\det e_\m^a$ and $R= R_{\m\n}^{\ ab} e_a^{\
\mu}e_b^{\ \nu}$. The constants $c_1,c_2$ and $c_3$ are arbitrary and, as we shall se later,
can be determined from additional conditions. 

The first term in the action (\ref{KomDejstvo}) is the topological Gauss-Bonnet term. It does not
contribute to the equations of motion and we shall not consider it further. The other two terms are
the Einstein-Hilbert term and the cosmological constant term. Note that, depending on the sign of
the coefficient $c_1+ 2c_2 + 2c_3$, the cosmological constant can be positive, negative or zero.
The spin connection $\omega_\mu^{ab}$ and the vierbeins $e_\mu^a$ are independent fields in the
model. Varying the action (\ref{KomDejstvo}) with respect to these fields gives 
two equations of
motion. The variation with respect to $\omega_\mu^{ab}$ gives the vanishing torsion equation. This
equation enables to express the spin connection in terms of the vierbeins. The variation of the
action (\ref{KomDejstvo}) with respect to $e_\mu^a$ gives the Einstein equation, written in terms of
the vierbiens. Finally, we conclude that after the symmetry breaking and solving for the spin
connection in terms of vierbeins we obtain the usual GR: Einstein equations with an arbitrary
cosmological constant. 

\vspace*{0.2cm}
{\it Noncommutative model and the low energy expansion:} We wish to generalize the
commutative model (\ref{KomDejstvo}) to the noncommutative space-time. We work in the canonically
deformed ($\theta$-constant noncommutative) space-time and use the approach of the deformation
quantization. The noncommutative fields are functions of commuting coordinates while their
multiplication is
given by the noncommutative (but associative) $\star$-product. In the case of the $\theta$-constant
NC space-time, the $\star$-product is the Moyal-Weyl product given by
\begin{eqnarray}
f\star g &=& e^{\frac{i}{2}\,\theta^{\a\b}\frac{\pa}{\pa x^\a}\frac{\pa}{ \pa
y^\b}} f (x) g (y)|_{y\to x}\nn\\
&=& f\cdot g + \frac{i}{2}\theta^{\a\b}(\pa_\a
f)(\pa_\b g)\nn\\
&& - \frac{1}{8}\theta^{\a\b}\theta^{\k\l}(\pa_\a\pa_\k
f)(\pa_\b\pa_\l g) +\dots .\label{MWStarProduct} 
\end{eqnarray}
The deformation parameter is a constant antisymmetric matrix $\theta^{\a\b}$. The $\star$-product
reduces to the usual commutative multiplication in the zeroth order in deformation parameter
and has higher order corrections that lead to $f\star g\neq g\star f$. 

The concept of gauge symmetry can be generalized to the NC setting. We shall use the enveloping
algebra approach and the Seiberg-Witten map \cite{SWMapEnvAlgebra}. The noncommutative
$SO(2,3)_\star$ gauge
transformation of the NC gauge field $\hat{\omega}_\mu$ is defined as
\begin{equation}
\delta_\epsilon^\star{\hat\omega}_\m = \pa_\m{\hat\Lambda}_\epsilon
+ i[ {\hat\Lambda}_\epsilon\ds {\hat\omega}_\m] ,\label{SO23Omega}
\end{equation}
with the NC gauge parameter $\hat{\Lambda}_\epsilon$. It can be shown that the algebra of NC
gauge transformations closes in the enveloping algebra only\footnote{Only in the case of $U(N)$ in
the fundamental representation the NC gauge transformation still close in the corresponding
Lie algebra.}. Since the enveloping algebra is infinite dimensional, it would seem that the NC gauge
theory should have infinitely many degrees of freedom. The idea of the Seiberg-Witten map is that
the NC gauge transformations are induced
by the commutative gauge transformations, $\delta_\epsilon \to
\delta^\star_\epsilon$. Then
\begin{equation}
\hat{\omega}_\m = \hat{\omega}_\m(\omega_\m),
\quad
\hat{\phi} = \hat{\phi}(\phi, \omega_\m) \, ,\nn
\end{equation}
that is, the NC fields are functions of the corresponding commutative fields. The method for
solving the SW map equations is introduced in \cite{kayahan,PC11} and the explicit solutions for
the
$SO(2,3)$ gauge group are given in \cite{MDVR-14}. The important result of this approach is
that there are no new degrees of freedom but new interaction terms appear. The
relations between
different (new) interaction terms are fixed by the SW map. 

The NC generalization of (\ref{KomDejstvo}) is given by
\begin{equation}
S_{NC} = c_1S_{1NC} + c_2S_{2NC} + c_3S_{3NC} \, ,\label{NCDejstvo}
\end{equation}
with
\begin{eqnarray}
S_{1NC} &=& \frac{il}{64\pi G_N}\tr \int{\rm d}^4x \epsilon^{\mu\nu\rho\sigma}
\hat{F}_{\mu\nu}\star \hat{F}_{\rho\sigma}\star \hat{\phi}\, ,\nn\\ 
S_{2NC} &=& \frac{1}{128 \pi G_{N}l}\tr \nn\\
&& \int d^{4}x \epsilon^{\mu \nu 
\rho \sigma}\hat\phi\star\hat F_{\mu \nu}\star\hat D_{\rho}\hat\phi\star\hat 
D_{\sigma}\hat\phi + c.c. \, , \nn \\
S_{3NC} &=& -\frac{i}{128 \pi G_{N}l}\tr\nn\\
&&\int \mathrm{d}^{4} x\, \varepsilon^{\mu\nu\rho\sigma}
D_\m 
\hat\phi \star D_\n \hat\phi\star
\hat{D}_{\rho}\hat{\phi} \star \hat{D}_{\sigma}\hat{\phi}\star \hat{\phi} \,
.\nn 
\end{eqnarray}
The action is invariant under the NC
$SO(2,3)_\star$ gauge symmetry and the SW map
guarantees that after the expansion it will be invariant under the commutative $SO(2,3)$
gauge symmetry.
Using the SW map solutions for the fields $\hat{F}_{\mu\nu}$ and $\hat{\phi}$ \cite{MDVR-14} and the
$\star$-product (\ref{MWStarProduct}), 
the action (\ref{NCDejstvo}) is expanded in orders of the NC parameter $\theta^{\alpha\beta}$. The
expansion is done around the commutative vacuum $\phi^5=l$, $\phi^a=0$, that is the symmetry
breaking is done after the expansion of NC fields and $\star$-products. The zeroth order of the
expansion is just the commutative action (\ref{KomDejstvo}). The first order correction vanishes.
This is an expected result: it was shown in \cite{SWmapApproach} that, if the NC gravity action is
real, then the first order (in the deformation parameter) correction has to vanish. This result
holds for a wide class of NC deformations, namely the deformations obtained by an Abelian
twist, see \cite{PL09}.

The second order correction is the first non-vanishing correction. The calculation is long,
some details and the result can be found in \cite{UsInPreparation}. The result, being long and
complicated is not so easy to analyze. Nevertheless, it enables to study different sectors of our
model: high curvature or low curvature, big, small or vanishing cosmological constant, etc. In this
Letter we analyze the low energy sector of the model. Therefore, in the expanded
NC gravity action we keep only terms that are of zeroth, first and second order 
in the derivatives of the vierbeins $e_\m^a$. The resulting action is given by
\begin{eqnarray} 
S_{NC} &=&
-\frac{1}{16\pi G_{N}}\int 
{\rm d}^{4}x\, \sqrt{-g}\Big(R-\frac{6}{l^2}(1+c_2+2c_3)\Big)\nn\\
&& +\frac{\theta^{\alpha\beta}\theta^{\gamma\delta}}{128\pi G_Nl^4}\int {\rm d}^{4} x\,  e
\Big( 
\frac{6+28c_2+56c_3}{l^{2}} g_{\alpha\gamma} g_{\beta\delta}\nn\\&&+(-2+15c_2+38c_3)
R_{\alpha\beta\gamma\delta} \nn\\&&
+(4-18c_2-44c_3)R_{\alpha\gamma\beta\delta} \nn\\
&&-(6+22c_2+36c_3)g_{\beta\delta}{R_{\alpha\mu\gamma}}^{\mu}\nn\\&& +(7-\frac{13}{2}c_2-7c_3)T_{\a\b}^aT_{\g\d a}\nn\\
&& +(-8+9c_2+18c_3)T_{\a\b\g}e^\m_a\nabla_\d e_\m^a 
+\dots   \Big) . 
\label{action-linearnoR}
\end{eqnarray}
$"\dots"$ include terms of the form: $\theta\theta \nabla e\nabla e$, 
$\theta\theta T\nabla e$, $\theta\theta \nabla T$ and  $\theta\theta T^2$, where $T$ labels the
torsion tensor $T_{\mu\nu}^a$. 
These terms are of no importance for the specific solution we analyze in this 
Letter. Therefore, we did not write them explicitly. The complete result is 
written in \cite{UsInPreparation}. It is quite interesting to observe that
$\theta^2 R$ term vanishes in the action 
(\ref{action-linearnoR}). Thus Einstein-Hilbet term is stable under NC 
corrections.    

The action (\ref{action-linearnoR}) is invariant under the local $SO(1,3)$ 
symmetry. The diffeomorphism symmetry is
broken, as expected. Terms that explicitly break this symmetry are of the form
$\theta^{\alpha\beta}\theta^{\gamma\delta}R_{\alpha\gamma\beta\delta}$ and
$\nabla_{\mu}e_{\rho}^{a}$. The first type of terms, despite the contracted indices, are not
scalars since $\theta^{\alpha\beta}$ is not a tensor but a constant matrix. The second type of
terms can be rewritten using the metricity condition as $\nabla_\mu e_\rho^{\ a} =
\Gamma_{\mu\rho}^\sigma e_\sigma^{\ a}$ with the affine connection $\Gamma_{\mu\rho}^\sigma$
explicitly appearing in the action.  

To obtain the equations of motion, we have to vary the action (\ref{action-linearnoR}) with respect
to the spin connection
$\omega_\mu^{ab}$ and the vierbein $e_{\mu}^{a}$. The equations are given by:
\begin{eqnarray}
&&R^{\n\m}-\frac12g^{\m\n}R+\L g^{\m\n}=\tau^{\m\n} 
,\label{DeltaE} \\
&&\tau^{\m\n} = \frac{e_a^\nu}{2e} \frac{\delta S^{(2)}_{NC}}{\delta e_\mu^a}\, ,\nn
\end{eqnarray}
\begin{eqnarray}
&&T_{bc}^{\ \ c}e_a^\m-T_{ac}^{\ \ c}e_b^\m+T_{ab}^{\ \ \m}=S_{ab}^{\ \ \m}\, ,\label{DeltaOmega}\\
&&S_{ab}^{\ \ \m} =  \frac{1}{2e}\frac{\delta S^{(2)}_{NC}}{\delta \omega_\mu^{ab}}\, ,\nn 
\end{eqnarray}
where $S^{(2)}_{NC}$ is the second order expansion of the action (\ref{action-linearnoR}). The
explicit expressions for $\tau^{\m\n}$ and $S_{ab}^{\ \ \m}$ are given in \cite{UsInPreparation}.
Even without explicitly solving these equations we can make some important 
comments. Looking at the
equation (\ref{DeltaOmega}) we see that noncommutativity is a source of torsion. Namely, even if
we
start from a torsion-free solution (in the zeroth order in the deformation parameter), NC
corrections will introduce non-zero torsion in the second order in the deformation parameter. The
analogous conclusion follows from the equation (\ref{DeltaE}): the flat space-time will become
curved due to the presence of noncommutativity. In the last part of the letter, we analyze one flat
space-time solution, namely the Minkowski space-time.

\vspace*{0.2cm}
{\it NC corrections to the Minkowski space-time:}
Minkowski space-time is a solution to the vacuum Einstein equations without the cosmological
constant.
Therefore, we first have to assume that $1+c_2+2c_3=0$, that is that the cosmological constant in
not present in the
zeroth order. Note that in our previous work \cite{MiAdSGrav, MDVR-14} we were not able to choose
the value of the cosmological constant, since we only worked with the action $S_{1NC}$. Adding the
other two actions $S_{2NC}$ and $S_{3NC}$ with arbitrary constants 
$c_1,c_2,c_3$ enables us to
study a wider class of NC gravity solutions.

Next we assume that the NC metric is of the form:
\begin{equation} 
g_{\m\n}=\eta_{\m\n}+h_{\m\n} ,\nn
\end{equation}
where $h_{\m\n}$ is a small correction that is second order in the deformation parameter
$\theta^{\alpha\beta}$. Inserting this ansatz into the action 
(\ref{action-linearnoR}) and varying
with respect to
$h_{\mu\nu}$ leads to the following equations of motion:
\begin{eqnarray}  
&&\frac{1}{2}(\pa_\s\pa^\n h^{\s\m}+\pa_\s\pa^\m h^{\s\n}-\pa^\m\pa^\n h -\Box 
h^{\m\n})\nn\\
&&-\frac12\eta^{\m\n}(\pa_\a\pa_\b h^{\a\b}-\Box h)\nn\\
&=&\frac{11}{4l^6}(2\eta_{\a\g}\theta^{\a\m}\theta^{\g\n}+\frac{1}{2}\eta_{\a\g}
\eta_{\b\d} \eta^{\m\n}\theta^{\a\b}\theta^{\g\d}) \, . 
\label{NCMinkEoMmetric} 
\end{eqnarray}
The right hand side of equation (\ref{NCMinkEoMmetric}) is constant. Therefore, these
inhomogeneous equations
are solved by a
general $h_{\mu\nu}$ quadratic in coordinates. We find solution of
the form:
\begin{eqnarray}
g_{00}&=& 1 - \frac{11}{2l^6}\theta^{0m}\theta^{0n}x^m 
x^n-\frac{11}{8l^6}\theta^{\a\b}\theta_{\a\b}r^2\nn\\
g^{0i} &=&  -\frac{11}{3l^6}\theta^{im}\theta^{0n}x^m 
x^n ,\nn\\ 
g_{ij}&=& -\d_{ij} -\frac{11}{6l^6}\theta^{im}\theta^{jn} 
x^mx^n\nn\\
&& +\frac{11}{24l^6}\d^{ij}\theta^{\a\b}\theta_{\a\b}r^2-\frac{11}{24l^6}\theta^{\a\b}
\theta_{\a\b}x^i x^j \, . \label{NCMinkowskiMetric}
\end{eqnarray}
The scalar curvature of this solution is given by
\begin{equation}
R=-\frac{11}{l^6}\theta^{\a\b}\theta^{\g\d}\eta_{\a\g}\eta_{\b\d} =const. 
\label{RNCMikowski}
\end{equation}
This shows that the noncommutativity induces curvature and the Minkowski space-time becomes
(A)dS-like. The sign of the scalar curvature will depend on the particular values of the
parameter
$\theta^{\a\b}$. The induced curvature is very small, being quadratic in $\theta^{\a\b}$ and it
will be difficult to measure it. However, qualitatively we showed that noncommutativity is a
source of curvature, just like matter or cosmological constant. It is tempting to try to relate the
quantity $\theta^{\a\b}\theta^{\g\d}\eta_{\a\g}\eta_{\b\d}/l^6$ with the 
actual value of the
cosmological
constant,
but to be able to to that we have to first study cosmological solutions and their corrections
induced by our NC gravity model.

The Reimann tensor for the solution (\ref{NCMinkowskiMetric}) can be calculated easily. A
very interesting and unexpected
observation follows: knowing the components of the Riemann tensor, the components of the metric
tensor in (\ref{NCMinkowskiMetric}) can be written as
\begin{eqnarray}
g_{00}&=&1-R_{0m0n}x^mx^n\, ,\nn\\ 
g_{0i}&=&-\frac23R_{0min}x^mx^n \, ,\nn\\ 
g_{ij}&=&-\d_{ij}-\frac13R_{imjn}x^mx^n \, .\label{FermiNCMinkowski}
\end{eqnarray}
This form of the metric tensor is typical for a special type of coordinates, the Fermi
normal coordinates. These coordinates are inertial coordinates of a local observer moving along a
geodesic. The time coordinate is just the proper time of the observer moving along the geodesic.
The space coordinates $x^i$ are defined as affine parameters along the geodesics 
in the hypersurface
orthogonal the actual geodesic of the observer. Unlike Riemann normal coordinates which can be
constructed in
a small neighborhood of a point, Fermi normal coordinates can
be constructed in a small neighborhood of a geodesic, that is inside a small cylinder
surrounding the geodesic \cite{FermiCoordiantes}. Along the geodesic these 
coordinates are inertial,
that is 
\begin{equation}
g_{\mu\nu}|_{geod.} = \eta_{\mu\nu}, \quad  \partial_\rho g_{\mu\nu}|_{geod.} = 0 \, . \nn
\end{equation}

The measurements performed by
the local observer moving along the geodesic are described in the Fermi normal coordinates.
Especially, she/he is the one that
measures $\theta^{\alpha\beta}$ to be constant! In any other reference frame (any other coordinate
system) observers will
measure $\theta^{\a\b}$ different from constant. 

With this observation we now understand the breaking of diffeomorphism symmetry in
the following way: there is a preferred
reference system defined by the Fermi normal coordinates and the NC parameter $\theta^{\a\b}$
is constant in that particular reference system. This suggests that the NC gravity should be
analyzed in the Fermi normal coordinates and that this choice of coordinates
could simplify the definitions and the analysis of the NC gravity
models in general. In our future work we plan to investigate other solutions of our NC gravity
model, like NC Schwartzschild solution and cosmological solutions. Especially, we are interested
in the emergence of Fermi normal
coordinates in these solutions and in this way we hope to gain a better understanding of NC gravity.

\vskip1cm \noindent
{\bf Acknowledgement}
\hskip0.3cm
We would like to thank Ilija Simonovi\'c, Milutin Blagojevi\' c,  Maja Buri\' c, Dragoljub Gocanin
and Nikola Konjik for
fruitful discussion and useful comments. The work is
supported by project
171031 of the Serbian Ministry of Education and Science and partially by
ICTP-SEENET-MTP
Project PRJ09 ”Cosmology and Strings” in frame of the Southeastern European
Network in Theoretical and Mathematical Physics.
%%%%%%%%%%%%%%%%%%%%%%%


\begin{thebibliography}{99}

\bibitem{GRWaves}
B. P. Abbott et al. (LIGO Scientific Collaboration and Virgo Collaboration), {\it Observation of
Gravitational Waves from a Binary Black Hole Merger}, Phys.\ Rev.\ Lett.\ {\bf 116}, 061102
(2016), [arXiv:1602.03837].

B. P. Abbott et al. (LIGO Scientific Collaboration and Virgo Collaboration), {\it GW151226:
Observation of Gravitational Waves from a 22-Solar-Mass Binary Black Hole Coalescence}
Phys.\ Rev.\ Lett. {\bf 116}, 241103 (2016).

\bibitem{DFR} 
S. Doplicher, K. Fredenhagen, J. E. Roberts,
{\it The Quantum structure of space-time at the Planck scale and quantum fields},
Commun.\ Math.\ Phys.\ {\bf 172}, 187-220 (1995), [hep-th/0303037]. 

\bibitem{TwistApproach}
P.Aschieri, C. Blohmann, M. Dimitrijevi\' c, F. Meyer, P. Schupp
and J. Wess, {\it A Gravity Theory on Noncommutative Spaces},
Class.\ Quant.\ Grav. {\bf 22}, 3511 (2005), [hep-th/0504183 ].

P. Aschieri, M. Dimitrijevi\' c, F. Meyer and  J. Wess,
{\it Noncommutative Geometry and Gravity}, Class.\ Quant.\ Grav. {\bf
23}, 1883 (2006), [hep-th/0510059].

T. Ohl and A. Schenckel, {\it  	
Cosmological and Black Hole Spacetimes in Twisted Noncommutative Gravity}, JHEP {\bf 0910} (2009)
052, [arXiv: 0906.2730].

\bibitem{EmGravityApproach} 
H. S. Yang, {\it Emergent gravity from noncommutative spacetime}, Int.\ J.\
Mod.\ Phys. {\bf A24}, 4473 (2009), [hep-th/0611174].

H. Steinacker, {\it Emergent Geometry and Gravity from Matrix Models: an
Introduction}, Class.\ Quant.\ Grav.\ {\bf 27}, 133001
(2010), [arXiv:1003.4134].

\bibitem{OtherApproaches}
M. Buri\' c and J. Madore, {\it Spherically Symmetric Noncommutative Space: $d =
4$}, Eur.\ Phys.\ J. {\bf C58},  347 (2008), [arXiv: 0807.0960].

M. Buri\' c and J. Madore, {\it On noncommutative spherically symmetric spaces},
arXiv:1401.3652.

L. Tomassini, S. Viaggiu, {\it Building non-commutative spacetimes at the Planck length for
Friedmann flat cosmologies}, Class.\ Quant.\ Grav. {\bf 31} 185001
(2014), [arXiv:1308.2767].

\bibitem{SWmapApproach} A. H. Chamseeddine, {\it Deforming Einstein's gravity},
Phys.\ Lett.\ B {\bf 504} 33 (2001), [hep-th/0009153].

M. A. Cardella and D. Zanon,
{\it Noncommutative deformation of four-dimensional gravity}, Class.\ Quant.\
Grav. {\bf 20}, L95 (2003), [hep-th/0212071].

\bibitem{PL09}
P. Aschieri and L. Castellani, {\it Noncommutative $D=4$ gravity
coupled to fermions} JHEP, {\bf 0906}, 086 (2009), [arXiv:0902.3823].

\bibitem{MajaJohn}
M. Buric, T. Grammatikopoulos, J. Madore, G. Zoupanos, {\it Gravity and the Structure of
Noncommutative Algebras}, JHEP {\bf 0604} 054, 2006, [hep-th/0603044].

M. Buric, J. Madore, G. Zoupanos, {\it The Energy-momentum of a Poisson structure},
Eur.\ Phys.\ J.\ {\bf C 55} 489-498, 2008, [arXiv:0709.3159].

\bibitem{stelle-west} K. S. Stelle and P. C. West, {\it Spontaneously
broken de Sitter symmetry and the gravitational holonomy group}, Phys.\ Rev D
{\bf 21}, 1466 (1980).

S. W. MacDowell and F. Mansouri,
{\it Unified geometrical theory of gravity and supergravity}, Phys.\ Rev.\ Lett.
{\bf 38}, 739 (1977).

P. K. Towsend, {\it Small-scale structure of spacetime as
the origin of the gravitation constant}, Phys.\ Rev.\ D {\bf 15},  2795 (1977).

\bibitem{UsInPreparation}
M. Dimitrijevi\' c \'Ciri\'c, B. Nikoli\'c and  V. Radovanovi\' c, in preparation.

\bibitem{Wilczek}
F. Wilczek, {\it  	
Riemann-Einstein structure from volume and gauge symmetry}, Phys.\ Rev.\ Lett.\ {\bf 80} (1998)
4851-4854, [hep-th/9801184].

\bibitem{SWMapEnvAlgebra}
B. Jur\v{c}o, L. M\"oller, S.~Schraml, P.~Schupp and J.~Wess,
{\it Construction of non-Abelian gauge theories on noncommutative spaces},
Eur.\ Phys.\ J.\ C{\bf 21}, 383 (2001), [hep-th/0104153].

N.~Seiberg and E.~Witten,
{\it String theory and noncommutative geometry},
JHEP {\bf 09}, 032 (1999), [hep-th/9908142].

\bibitem{kayahan} K. Ulker and B. Yapiskan, {\it Seiberg-Witten maps to all
orders}, Phys.\ Rev.\ D {\bf 77}, 065006 (2008), [arXiv: 0712.0506].

\bibitem{PC11} P. Aschieri and L. Castellani, {\it Noncommutative gravity 
coupled to fermions: second order expansion via Seiberg-Witten map},  JHEP {\bf
1207} 184 (2012), [arXiv:1111.4822].

P. Aschieri, L. Castellani and M. Dimitrijevi\'c, {\it
Noncommutative gravity at second order
via Seiberg-Witten map}, Phys. Rev. D {\bf 87},  024017 (2013),
[arXiv:1207.4346]

\bibitem{MDVR-14}
M. Dimitrijevi\' c and  V. Radovanovi\' c,
{\it Noncommutative $SO(2,3) $ gauge theory and noncommutative gravity},
Phys. Rev. D {\bf 89}, 125021 (2014), [arXiv:1404.4213].

\bibitem{MiAdSGrav}
M. Dimitrijevi\' c, V. Radovanovi\' c and H. \v Stefan\v ci\' c,
{\it AdS-inspired noncommutative gravity on the Moyal plane},
Phys. Rev. D {\bf 86}, 105041 (2012), [arXiv:1207.4675].

\bibitem{FermiCoordiantes}
F.K. Manasse and C.W. Misner, {\it Fermi Normal Coordinates and Some Basic Concepts in
Differential Geometry}, J.\ Math.\ Phys.\ {\bf 4} (1963) 735-745.

C. Chicone and B. Mashoon, {\it Explicit Fermi coordinates and tidal dynamics in de Sitter and Godel
spacetimes}, Phys.\ Rev.\ {\bf D 74} (2006) 064019, [gr-qc/0511129]. 

D. Klein and E. Randles, {\it  	
Fermi coordinates, simultaneity, and expanding space in Robertson-Walker cosmologies}, Annales Henri
Poincare 12 (2011) 303-328, [arXiv:1010.0588].


\end{thebibliography}
\end{document}